\documentclass[useAMS,usenatbib]{mn2e}
\usepackage{amssymb,amsmath,epsfig,times}
\def\ev{{\rm\thinspace eV}}
\def\kev{{\rm\thinspace keV}}

\def\mpc{{\rm\thinspace Mpc}}

\def\kmps{{\rm\thinspace km \thinspace s^{-1}}}

\def\crexp{{\rm\thinspace km^{2} \thinspace sr \thinspace yr}}

\def\ergcms{{\rm\thinspace erg \thinspace cm^{-2} \thinspace s^{-1}}}

\title[AGN as Sources of UHECRs]{On Active Galactic Nuclei as Sources
  of Ultra-High Energy Cosmic Rays}  
\author[M. R. George et al.]{M. R. George,$^1$\thanks{Email:
    mrg@ast.cam.ac.uk} A. C. Fabian,$^1$ W. H. Baumgartner,$^{2,3}$
  R. F. Mushotzky,$^2$ and J. Tueller$^2$ \\
  $^1$Institute of Astronomy, Madingley Road, Cambridge CB3 0HA \\
  $^2$NASA/Goddard Space Flight Center, Code 662, Greenbelt, MD 20771,
  USA \\
  $^3$University of Maryland, Baltimore County, 1000 Hilltop Circle,
  Baltimore, MD 21250, USA
}
\date{Submitted 3 April 2008; Accepted 14 May 2008}

\begin{document}  

\maketitle

\begin{abstract}
We measure the correlation between sky coordinates of the \emph{Swift} BAT
catalogue of active galactic nuclei with the arrival directions of the
highest energy cosmic rays detected by the Auger Observatory. The
statistically complete, hard X-ray catalogue helps to distinguish
between AGN and other source candidates that follow the distribution
of local large-scale structure. The positions of the full catalogue
are marginally uncorrelated with the cosmic ray arrival directions,
but when weighted by their hard X-ray flux, AGN within $100\mpc$ are
correlated at a significance level of $98$ per~cent. This correlation
sharply decreases for sources beyond $\sim{}100\mpc$, suggestive of a
GZK suppression. We discuss the implications for determining the
mechanism that accelerates particles to these extreme energies in
excess of $10^{19}\ev$.
\end{abstract}


\section{Introduction}
The recent announcement of anisotropy in the arrival directions of
ultra-high energy cosmic rays (UHECRs) by the Auger Collaboration
\citep{auger07a,auger07b} has yielded much interest. By correlating
the UHECR arrival directions with the positons of nearby active
galactic nuclei (AGN) on the sky, they suggest that AGN are
responsible for accelerating these particles.

Modern experiments are approaching the size and sophistication needed
to measure the energy and direction of UHECRs with enough precision to
reliably determine their sources. The Auger Observatory combines
measurements of {\v C}erenkov radiation from particle interactions in
surface detectors and fluorescence from molecules in the atmosphere
excited by the cascade \citep{auger04}. For reviews of earlier
experimental work, see \citet{nagano00, cronin05, sokolsky07}. 

The origin of particles with energies
$\gtrsim\thinspace{}10^{19}\ev$ has been a longstanding mystery in
high energy astrophysics (see Hillas 1984 \nocite{hillas84} for a
review). Supernovae, gamma-ray bursts, pulsars, shock fronts in galaxy
clusters, decays of exotic massive particles, and various classes of
AGN have all been proposed as acceleration or generation sites. All
but the highest energy cosmic rays are thought to be significantly
deflected by Galactic magnetic fields, which makes it difficult to
trace arrival directions back to astronomical sources. Numerous
attempts have been made to correlate different source catalogues with
events observed by cosmic ray detectors including Yakutsk, Fly's Eye,
AGASA, HiRes, and now Auger \citep[\emph{e.g.,} ][]{stanev95, singh04,
  hague07, gorbunov07, kashti08, ivanov08}. Related studies have
analysed the auto-correlation of sources and events as well as the
occurrence of multiple co-located UHECR detections to place
constraints on the number of sources responsible for these events
\citep{takeda99, cuoco07}. AGN have long been known to be capable of
generating electric potential differences in excess of $10^{19}$~volts
\citep{burns82}, and given the further suggestion from the Auger team
that their positions are correlated with UHECR arrival directions, we
focus only on these objects. 

The correlation of Auger events with extragalactic sources given
by \citet{auger07b} is based on an inhomogeneous list of AGN
collected from the literature \citep[ hereafter V-C]{vc06}. As pointed
out by others \citep{kashti08} and explained in the companion paper to
the catalogue itself, the compilation is not complete and should not
be used for statistical purposes. While using such a list can be
adequate for disproving the hypothesis of an isotropic distribution of
arrival directions, positive proof of correlation with sources calls
for a statistically complete sample. Additionally, the distribution of
local AGN is similar to that for all local large-scale structure. A
complete sample of sources selected for a certain property can
distinguish between different tracers of large-scale structure and
also determine if a particular factor is relevant to UHECR acceleration.

In this letter, we study the correlation of Auger UHECR arrival
directions with the positions of AGN from the hard X-ray selected
\emph{Swift} BAT catalogue \citep{tueller07}. Selection from the hard X-ray
band reduces the bias due to absorption that impacts lower energy
bands. Optically selected lists, including many that are compiled into
the V-C catalogue, are less likely to include the obscured nuclei that
make up an increasing fraction of the AGN population at low
redshifts \citep{brandt05}. \citet{matt00} deduce from hard
X-ray observations that heavily obscured sources must locally
outnumber unobscured sources by a large factor; the nearest 3 AGN are
all highly obscured.

Since hard X-ray observations are less sensitive to obscuration, they
provide a more accurate indicator of the accretion rate and the
intrinsic luminosity of an AGN. We aim to test the idea that the intrinsic
luminosity of an AGN is a sign of its particle accelerating ability
by measuring the relation between hard X-ray flux and the positional
correlation of AGN with UHECR arrival directions. In the next section, we
will present the catalogues of UHECRs and AGN. \S\ref{analysis} will
describe the statistical tests used to measure correlation between the
two lists as well as the results of these tests. We discuss the
implications of these results and compare them to previous analyses in
\S\ref{discussion}, and \S\ref{conclusion} concludes the paper. 

\section{Data}
In this section we discuss the catalogues of UHECRs and AGN used for
testing the correlation between events and sources. We do not
combine data from different cosmic ray observatories because of
possible inconsistencies in energy calibration and angular
resolution. A statistically complete, hard X-ray selected sample of
AGN will reduce the bias toward unabsorbed sources present in the V-C
catalogue, which has mainly been selected from optical
observations. This sample should clarify whether the accretion
processes in AGN that generate hard X-rays are relevant to high energy
particle acceleration. By narrowing the sample from a large and
inhomogeneous list of AGN, we can test the possibility that
the observed correlation of UHECRs and AGN is merely due to the fact
that AGN are distributed similarly to other sources in the local
structure.

\subsection{UHECR events}
The Auger Observatory \citep{auger04,auger07b} has been operating stably in
Argentina since 2004, using a hybrid system of telescopes to measure
fluorescence in the atmosphere and water tanks to detect {\v C}erenkov
light from relativistic particles. The available UHECR list is
comprised of 27 events with energies above
$5.7\times{}10^{19}\ev$ from an integrated exposure of $9000\crexp$.

The relative exposure is independent of energy in this range,
nearly uniform in right ascension, and has a declination dependence
given by \citet{sommers01}. The latitude of the Auger Observatory is
$-35.2\degr$ and it has a maximum zenith angle acceptance of $60\degr$.

The event arrival directions are determined with an angular
resolution of better than $1\degr$. However, magnetic fields of
unknown strength will deflect charged particles on their
trajectories through space. The advantage of studying the highest
energy events is that this deflection is minimized, but it can still
be up to $\sim{}10\degr$ in the Galactic field \citep[\emph{e.g.},
][]{stanev97}. Magnetohydrodynamical simulations of extragalactic
fields have produced conflicting predictions for UHECR deflections,
ranging from tens of degrees \citep{sigl04} to less than a few degrees
\citep{dolag05}.

\subsection{AGN catalogue}
We compare the arrival directions of Auger UHECRs with the locations
of AGN in the \emph{SWIFT} BAT survey. Details of the survey and
results from the first 3 and 9 months of observations are presented by
\citet{markwardt05} and \citet{tueller07}; we use a catalogue
  updated with the first 22 months of data.

The list includes 254 objects that have known redshifts and signal to
noise ratios greater than 4.8, corresponding to one false detection
across the sky. Source identification is incomplete at low
  Galactic latitudes ($|b|<15\degr$), but the hard X-ray coverage
($14-195\kev$) provides a uniquely unbiased sample of local AGN over
the rest of the sky. The median astrometric uncertainty is 1.7 arcmin
and the limiting sensitivity is a few times $10^{-11}\ergcms$. All but
four sources in the 22 month catalogue have X-ray positions and
spectra from previous observations or follow-up with the \emph{SWIFT}~XRT.

We restrict our statistical analysis to the complete
sample within the Auger field of view ($\rm{S/N}>4.8,$ $|b|>15\degr,
\delta<24.8\degr$), leaving 138 AGN in the catalogue. For comparison
with the original correlation between Auger events and AGN, there are
57 \emph{Swift} sources within the Auger field of view at distances
less than $100\mpc$, and 6 of these are not in the V-C catalogue. The
V-C catalogue has 410 objects within $100\mpc$ and the Auger field of view
outside of the Galactic plane. Throughout this paper, distances are
determined using a cosmology of $\rm{H}_0=71\kmps\mpc^{-1}$ and
$\Omega_m=0.27$ with a flat universe. 

\section{Analysis}
\label{analysis}
\begin{figure*}
  \begin{center}
    \leavevmode
      \epsfxsize=14.5cm
      \epsfbox{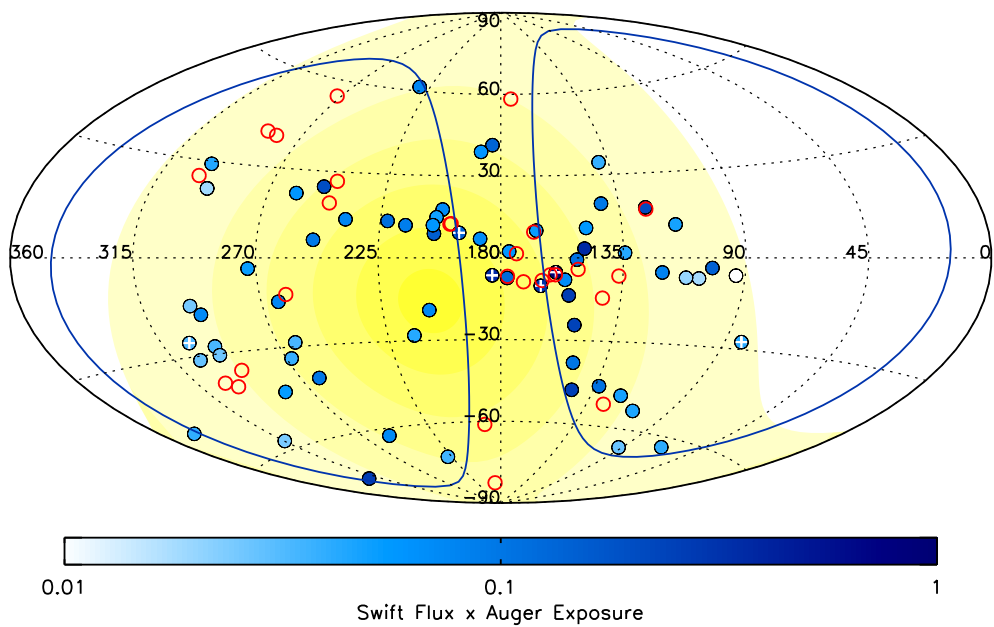}
       \caption{Map of Auger UHECRs (open red circles) and BAT AGN
         within $100\mpc$ (filled blue circles) in supergalactic
         coordinates \citep{devaucouleurs76}. The blue colour depth is
         scaled by the hard X-ray flux and Auger exposure, relative to
         Cen~A. The 6 AGN in the catalogue within $20\mpc$ are marked
         with white crosses, with Cen~A at
         $(159.7\degr,-5.2\degr)$. Yellow contours have equal
         integrated exposures. Blue boundaries show where the AGN
         catalgoue is incompete due to the Galactic plane, $|b|<15\degr$.}
     \label{map}
  \end{center}
\end{figure*}

Figure~\ref{map} shows the locations of Auger UHECRs and AGN from the
BAT catalogue in supergalactic coordinates. The relative exposure of
Auger observations is plotted in yellow contours, and the AGN are
shaded blue according to the product of their X-ray flux and the
relative Auger observation time. Blue curves indicate where the BAT
catalogue is incomplete at low Galactic latitudes, $|b|<15\degr$.

%

Visually, we can see several interesting correlations and
non-correlations. Cen~A is the brightest source and has two Auger
events within $3\degr$. The likelihood of observing 2 of the 27 events within
this angular separation of the brightest AGN in the entire catalogue for an
isotropic distribution of arrival directions in Auger's field of view
is about $3\times 10^{-4}$. The 6 AGN within $20\mpc$ are marked
with white crosses, and 4 of these (including Cen~A) have UHECR
arrival directions within $6\degr$. If this proportion of observed
UHECRs does in fact arrive from nearby AGN and the sources are roughly
isotropic on large scales, we can infer that there must be significant
propagation losses for cosmic rays from distant sources. However,
clusters of 2 and 3 Auger events have no corresponding AGN, and the
overall correlation is unclear. Certainly there is not a 1-to-1
correspondence between all of the observed UHECRs and a hard X-ray
emitting AGN.

To quantify the likelihood that the two sets of locations arise from
the same population of sources, we employ the two-dimensional
generalization of the K-S test \citep{fasano87,press07}. This test can
be used to compare a data sample against a model distribution, or to
compare the properties of two data samples against each other. To test
the null hypothesis that a given class of sources is responsible for
UHECRs, one could construct a model of the expected cosmic ray flux
map using a catalogue of positions, intrinsic luminosities, distances,
expected absorption losses, and the detector acceptance as a function
of position. Additionally, one should account for interactions with
background radiation which produce propagation losses known as the GZK
effect \citep{greisen66,zatsepin66}. \citet{singh04} used this
approach to measure the correlation of infrared galaxies with UHECR
events observed by AGASA, including models for the injection spectra
and propagation effects.

We might expect the hard X-ray flux, as a measure of the
intrinsic power of the source, to be a reasonable proxy for the
cosmic ray flux. But because a model of GZK effects requires an input
spectrum of cosmic rays and the injection mechanism is unknown, we opt
to begin by simply comparing the AGN locations with the UHECR arrival
directions. Subsequently, we weight these positions by the product of
the X-ray flux and the relative Auger exposure.

The two-dimensional K-S test measures the fraction of data points
lying in the natural quadrants defined around each data point. The
statistic $D$ is the maximum difference between the fractions of each
of the two data sets that lie in a quadrant, found after iterating over all
points and their respective quadrants. The strength of the correlation
between two catalogues is reported as the integral probability
distribution $P(D\sqrt{n}>\rm{observed})$, where $n=N_1N_2/(N_1+N_2)$,
and $N_1$ and $N_2$ are the number of data points in the two
sets. This measurement can be used to determine the similarity of sets
of positions on the sky. Statistically speaking, high values of $P$ do
not \emph{prove} the null hypothesis that the two catalogues arise
from the same population, but low values $(P\lesssim 20$~per~cent)
call for the rejection of that null hypothesis. 

The probability that the two sets of data are from the same population
can be determined with an analytical approximation if the number of
data points is sufficient, but due to the limited number of events
available and the irregular shape of the allowed coordinate plane, we
generate Monte Carlo simulations of cosmic ray arrival directions to
compare with the observed events. For each set of AGN considered,
10,000 lists of positions are randomly generated in a distribution on
the sky that is proportional to the Auger exposure, excluding low
galactic latitudes. Each list has the same number of UHECRs as
observed in the selected region, and has its $D$ statistic calculated
in the same way as the real data. The significance of the correlation
between a catalogue of AGN and the observed UHECRs is reported as
$P(D\sqrt{n}>\rm{observed})$, which is the percentage of randomly
generated UHECR lists that have a higher value of $D\sqrt{n}$ than the
real data under consideration. Thus, extreme values of $P$ suggest that
the observed events are not distributed isotropically, and high (low) values
indicate a good (poor) correlation between the Auger UHECRs and the given AGN
list.

It is important to ensure that the test does reliably
differentiate between a set of points correlated with a reference
catalogue and an isotropically distributed set. We picked random
subsets of AGN from the larger V-C catalogue to serve both as a list of AGN
and a list of UHECRs, with the number of sources and events equal to
those in the actual tests on \emph{Swift} and Auger data. Average
values for the correlation showed $P\gtrsim 70$~per~cent, setting a
rough threshold for accepting that two data sets belong to the same
distribution.

The test differs from the one performed by the Auger team in that the
angular distance between the source position and arrival direction is
not a parameter of the fit, so it is not as sensitive to the
undetermined size of magnetic deflections. The angular scale of  the
test's sensitivity is not clearly defined and depends on the
separation of points in the data set, but Monte Carlo simulations
indicate that correlations decrease monotonically with angular
separation, and that this decline has a scale length of approximately
$10-15\degr$. Additionally, we use the full energy range of the
published Auger UHECR catalogue, which has a threshold selected to
maximize the correlation in their analysis. We allow source distances
in the full range of the AGN catalogue, though completeness becomes an
issue at higher distances.

In our application, we use the equatorial positions for the Auger
events and \emph{Swift} AGN as the data points. Restricting the data
sets to regions and significance levels where the BAT catalogue is
complete $(|b|>15\degr, \rm{S/N}\geq 4.8)$ and to where the Auger
exposure is non-zero $(\delta<24.8\degr)$ leaves 138 AGN and 19
UHECRs. For these catalogues, we measure $P=50$ per~cent. This
value jumps to 98 per~cent after weighting the AGN coordinates by
their hard X-ray flux and relative Auger exposure. If we cut the 
AGN catalogue to those 57 with distances less than $100\mpc$, the
unweighted probability is 84 per~cent. The flux-weighted value is
unchanged, indicating that the correlation is dominated by nearby,
bright sources. However, the removal of Cen~A does not alter the
correlation values by more than a few per~cent. Cuts on the catalogue
to distances smaller than $100\mpc$ do increase the correlation
slightly, but quickly the number of sources becomes too small to
obtain reliable values from the test. For comparison, the 410 sources
in the V-C catalogue with $d<100\mpc$, $|b|>15\degr$, and
$\delta<24.8\degr$ produce a correlation with the UHECRs of only 55~per~cent.

\begin{figure}
    \leavevmode
      \epsfxsize=8cm
      \epsfbox{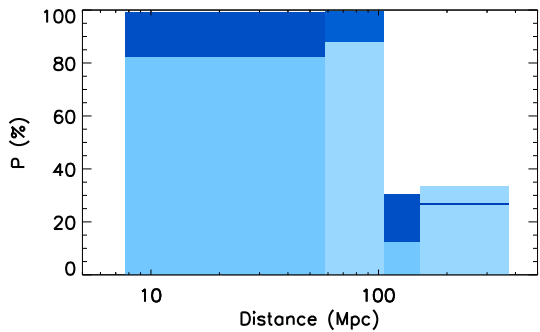}
       \caption{Redshift dependence of the correlation between AGN
         positons and UHECR arrival directions. Light (dark) boxes
         show the unweighted (flux-weighted) values of $P$, defined in
         the text. In the most distant bin, flux-weighting decreases
         the significance of the correlation. There are 30 AGN in each
         comoving distance bin.}
     \label{zdep}
\end{figure}

To determine the redshift dependence of the correlation, we test 4
redshift-sorted bins of 30 AGN each, with mean comoving distances of
35, 80, 131, and $248\mpc$. The results are plotted in
Figure~\ref{zdep}, which shows a clear decrease in the correlation at
larger distances. Replicating the plot with different bin sizes, we
consistently see a sharp decrease in the correlation near $100\mpc$.

We performed a similar test on the luminosity-dependence of the $P$
values and found that the correlation in the full catalogue is
actually lower for more luminous sources than for dimmer ones. It
is possible that the strong radiation field surrounding luminous
AGN interferes with UHECR propagation \citep{dermer07}. But when
restricting the AGN to distances less than $100\mpc$, the trend
reverses and more luminous sources tend to correlate better than less
luminous ones. In both cases, the most powerful AGN are at a
higher average distance than dim ones. We interpret these results to
imply that the cosmic ray flux is correlated with the luminosity of
AGN, but propagation losses on a distance scale of $\sim{}100\mpc$
interfere with the effect.

We would also like to test the correlation for subclasses of
AGN. There are 46 ``Type 1'' objects (Seyfert 1/1.2 galaxies) and 53 ``Type
2'' objects (Seyfert 1.8/1.9/2 galaxies) at high Galactic latitudes
and within the Auger exposure. There are not enough objects in other
subclasses such as blazars or quasars to obtain statistically
significant results. For Type 1 objects, the unweighted and weighted
$P$ values are 59 and 96 per~cent, respectively, while for Type 2
objects these values are 53 and 99 per~cent. For both types, excluding
sources at distances greater than $100\mpc$ increases the unweighted
correlation by $\sim$30 percentage points, while the flux-weighted
values remain the same.

The overall results are consistent with visual expectations, in the sense
that the correlation between the full catalogues is marginal at best,
but various cuts and weighting factors can improve the
correspondence. We risk falling into the trap of using
\emph{a~posteriori} cuts to find the best matches and reporting prejudical
results. Lengthy arguments have taken place over the usefulness of
applying ``penalty factors'' to account for the statistical biases
produced by these cuts \citep{evans03,finley04}, but we omit this type
of analysis and merely point out that the stated probabilities should
be taken with these caveats in mind.

\section{Discussion}
\label{discussion}
Our results from the K-S tests do not cause us to reject the null
hypothesis that AGN are the source of UHECRs, in disagreement with the
conclusion of \citet{gorbunov07}. On the contrary, we find significant
correlations that suggest that AGN are the sources of UHECRs,
supporting the claims of \citet{auger07a}. The hard X-ray catalogue
used in this paper produces a higher correlation with UHECR arrival
directions than the larger V-C catalogue which was previously
tested. From this information alone, it is still possible that another
type of source with a local distribution similar to that of AGN is
responsible for UHECR acceleration. But when weighting by the hard
X-ray flux or selecting the most luminous nearby sources, the measured
correlation between AGN and UHECRs increases. Thus, a confounding
class of sources would have to trace both the local spatial and
luminosity distributions of AGN, which seems unlikely.

If AGN are positively identified as the sources of UHECRs, the second
main result of this paper is the significant decline in correlation at
a distance of $\sim{}100\mpc$. This drop is predicted from the GZK cutoff, which
suggests that the flux of cosmic rays in the energy range considered
here is depleted due to photopion production and pair creation from
interactions with the cosmic microwave background. For
cosmic rays with energies above $5\times10^{19}\ev$, the
observable ``horizon'' is expected to be of order $100\mpc$, with
deviations due to the distribution of sources and the masses of
accelerated nuclei \citep{harari06}. HiRes has detected a suppression
of the cosmic ray spectrum at energies above $6\times10^{19}\ev$ which
has been attributed to the GZK cutoff \citep{abbasi08}. Though this
detection has been supported by the Auger data, it is conceivable that
the intrinsic spectrum of cosmic rays falls off at this energy due to
some cutoff in the acceleration mechanism. An observation that UHECRs
arrive from nearby sources and not distant ones would provide direct
evidence for the GZK effect. 

The drop observed in the correlation between AGN positions and UHECR
arrival directions at $100\mpc$, as shown in Figure~\ref{zdep}, is
suggestive of such a suppression. It may arise due to other
effects including increased angular deflections due to
extragalactic magnetic fields and incompleteness of the AGN catalogue
beyond this distance \citep{auger07b}. However, if magnetic deflections are the
culprit we would expect the correlation to begin decreasing before
$100\mpc$, which is not observed. Catalogue incompleteness affects the
detection of less luminous AGN at large distances, but for local AGN the
correlation reduces with faintness. Additionally, the evidence
discussed earlier that 4 nearby AGN have associated events does, from
Olber's paradox considerations, strongly argue that the sources of all
detected UHECRs are at very low redshift.

The association of two UHECRs with the brightest AGN in the
\emph{Swift} catalogue allows for speculation about the nature of
particle acceleration mechanisms. Several studies have suggested that
blazars are likely candidates for UHECR sources \citep[\emph{e.g.,} ][]{tinyakov01}, though the Auger data do not support a correlation
\citep{harari07}. Cen~A is a nearby jetted AGN with two large radio lobes
visible from the side. If Cen~A is in fact responsible for some of
the observed UHECRs, then cosmic rays do not need to be emitted
directly along a jet toward the observer. It is conceivable that jets
may still be important for cosmic ray acceleration and that the events
seen towards Cen~A were deflected by strong magnetic fields
surrounding the AGN. Our division of the AGN catalogue into ``Type 1''
and ``Type 2'' classes yielded similar results from the K-S test for
correlation, but further tests on distinct classes of AGN, including
more heavily obscured sources\footnote{We searched for a correlation
  with several heavily obscured sources too faint to be detected in
  the BAT catalogue, selected from \citet{levenson06}. We have not
  identified any obscured AGN located near the UHECR arrival
  directions, but the sample size is too small to draw conclusions
  about this population of AGN in relation to the UHECRs.}, will help
clarify which mechanism or orientation is necessary to produce the
observed UHECRs. 

As data for a larger number of UHECRs become available, we will be
able to further differentiate between the properties of acceleration
sites. We have shown that the correlation between AGN positions and
UHECR arrival directions does vary with the hard X-ray flux and the
distance to the source. It will be interesting to test other
properties to constrain models for the acceleration mechanism. Black
hole spin may be an important parameter but has not yet been tested
because of the difficulty in measuring its value for a sizable sample
of sources. Many models exist for the extraction of energy from black
holes \citep[\emph{e.g.,} ][]{blandford77}, and a further analysis of
the influence of the model parameters on the correlation between
source positions and UHECR directions will more precisely determine
which sources are accelerating these particles. Conversely, when the
cosmic ray sources are clearly identified, more detailed observations
in the realm of particle astronomy will improve our understanding of
the physical mechanisms taking place in these sources.

\section{Conclusion}
\label{conclusion}
We have measured the correlation between the positions of the
\emph{Swift} hard X-ray selected catalogue of AGN and the arrival
directions of UHECRs observed by Auger. No significant correlation is
found for the full catalogue, but imposing weights based on the
expected cosmic ray flux results in a correlation with a significance
level of 98 per~cent. A steep drop in the correlation for sources
beyond $100\mpc$ suggests that the GZK effect does suppress the cosmic
ray flux at these energy and distance scales.

Though our study lends support to the claims of the Auger
Collaboration that AGN are responsible for UHECRs, the connection should
still be tested with better statistics as observations continue. The
identification of 2 UHECRs from the direction of Cen~A and the
improvement of the correlation when weighting by hard X-ray flux offer
tantalizing hints about the acceleration mechanism. Though the nature
of these exotic UHECRs remains largely unknown, future studies in
particle astronomy may lend insight into other areas such as Galactic
and extragalactic magnetic fields and the physics behind AGN engines.

\section*{Acknowledgments}
We thank Annalisa Celotti for helpful discussions. MRG is supported by
a Herschel Smith Scholarship and ACF thanks the Royal Society for
support.

\bibliographystyle{mn2e}	
\bibliography{uhecrs20080513} 	

\end{document}